\documentclass[twocolumn,trackchanges]{aastex631}

\begin{document}

\title{[C/N] Ages and Extra-Mixing for [Fe/H] $<-$ 0.5: Insights from the LMC and SMC}

\newcommand{\logg}{log$g$}
\newcommand{\Msun}{$M_{\odot}$}
\newcommand{\MSU}{Montana State University, P.O. Box 173840, Bozeman, MT 59717-3840, USA}
\newcommand{\osu}{Department of Astronomy, The Ohio State University, Columbus, 140 W 18th Ave, OH 43210, USA}
\newcommand{\ccapp}{Center for Cosmology and Astroparticle Physics (CCAPP), The Ohio State University, 191 W. Woodruff Ave., Columbus, OH 43210, USA}

\author[0000-0003-4769-3273]{Yuxi(Lucy) Lu}
\affiliation{\osu}
\affiliation{\ccapp}

\author[0000-0002-2854-5796]{John D Roberts}
\affiliation{\osu}
\affiliation{\ccapp}

\author[0000-0002-6553-7082]{Joshua T. Povick}
\affiliation{\MSU}

\author[0000-0002-7549-7766]{Marc H. Pinsonneault}
\affiliation{\osu}
\affiliation{\ccapp}

\author[0000-0003-0929-6541]{Madeline Howell}
\affiliation{\osu}
\affiliation{\ccapp}

\author[0000-0002-1793-3689]{David L. Nidever}
\affiliation{\MSU}

\author[0000-0001-7258-1834]{Jennifer A. Johnson}
\affiliation{\osu}
\affiliation{\ccapp}



\begin{abstract}
The [C/N]–age relation has become a powerful tool for reconstructing the formation history of the Milky Way (MW), providing the largest age sample for field giant stars. 
However, at metallicities below [Fe/H] $< -0.5$, stellar surfaces are altered by a poorly understood process known as extra mixing, which modifies [C/N] in a mass- and metallicity-dependent manner. 
This effect complicates the application of the traditional [C/N]–age relation in metal-poor regimes. 
Within the MW, constraining the mass dependence of extra mixing is particularly challenging because stars at [Fe/H] $< -0.5$ are predominantly old and therefore low-mass, leading to strong degeneracies between mass and metallicity.
In this work, we explore the potential of the Magellanic Clouds (MCs) to disentangle these effects and constrain extra mixing as a function of age and metallicity. 
By comparing empirical corrections calibrated in the MW with predictions from thermohaline mixing models, we isolate the mass dependence of extra mixing in the MCs down to [Fe/H] $\sim-0.7$. 
We find that the empirical calibration performs well for lower-mass stars ($< 1.25$ \Msun), while theoretical models successfully reproduce the observed mass dependence down to $\sim$ 1.25 \Msun. 
We further present the first observational evidence that extra mixing becomes ineffective above $\sim$ 1.8 \Msun\ at [Fe/H] $\sim -0.7$.
Our results demonstrate the feasibility of deriving [C/N]-based ages for individual stars in external galaxies. 
Future observations targeting higher-\logg\ or fainter stars in the MCs will provide stronger constraints on extra-mixing processes and enable the calibration of [C/N]-age relation that can be applied to low-metallicity individual stars in the MW or external galaxies.

\end{abstract}

\keywords{Stellar physics(1621) --- Large Magellanic Cloud(903) --- Stellar ages(1581) --- Stellar abundances(1577) --- Giant stars(655)}


\section{Introduction} \label{sec:intro}
Stars mix material processed during the CNO-cycle to their surfaces as they ascend the lower giant branch during the first dredge-up phase \citep{Iben1967}, affecting their carbon-to-nitrogen ratio ([C/N]) in a way that is mass-dependent.
As a result, [C/N] is commonly used as an age indicator for giant stars, since it traces stellar mass, which correlates with age on the giant branch due to the shorter main-sequence lifetimes of more massive stars \citep[e.g.,][]{Martig2016, Ness2016, stonemartinez2025, Roberts2025}.
[C/N]-based ages are widely used in Galactic archaeology because these abundances are straightforward to measure, and empirical calibrations can be applied to distant stars with minimal sensitivity to dust extinction, which affects many other age-dating methods.
However, at low metallicity (about [Fe/H]$<-$0.5), giants undergo additional ``extra-mixing'' (also called the non-canonical mixing) after the first dredge-up that alters surface [C/N] in ways that are still poorly understood, making [C/N]-based ages uncalibrated in this regime \citep[e.g.,][]{Gratton2000}.

Thermohaline mixing \citep[e.g.,][]{Ulrich1972, Charbonnel2007, Charbonnel2010}, caused by the molecular weight inversion created by the $^3$He($^3$He, 2p)$^4$He reaction in the external wing of the hydrogen-burning shell, was proposed to explain the non-canonical mixing. 
Theory predicts that the effects of mixing are a strong function of mass, at solar metallicity, thermohaline mixing stops being effective at a mass of about 1.5--2.0 \Msun\ \citep[e.g.,][]{Cantiello2010, Charbonnel2010, Lagarde2012}.
However, thermohaline mixing as the extra-mixing mechanism on the Red Giant Branch (RGB) might not be efficient enough \citep[e.g.,][]{Denissenkov2010} and does not simultaneously match carbon and lithium abundances in globular clusters and field stars \citep[e.g.,][]{Angelou2015, Henkel2018, Tayar2022}.

Empirical calibration of extra-mixing as a function of metallicity has been done using high-$\alpha$ stars in the Milky Way (MW), assuming they are all of the same mass, between 0.9-1.1 \Msun\ \citep{Shetrone2019}.
This assumption, according to APOKASC--2 \citep{Pinsonneault2018} and the newest APOKASC--3 asteroseismic catalog \citep{Pinsonneault2025}, is valid.
As expected, they found the decrease in [C/N] from extra-mixing increases towards lower metallicity, and additional mixing continues towards lower \logg\ for stars with [Fe/H] $< -$1, also as seen in globular clusters \citep[e.g.,][]{Angelou2015}.

Extending the calibration to stars with a range of masses at low metallicity is difficult, as low-metallicity ([Fe/H] $< -$0.5) in the MW are all old, and thus, low-mass.
Moreover, a large fraction of field stars with [Fe/H] $< -$1, even after excluding known members from dwarf galaxies, could still be halo stars, with possibly different birth abundances.
As a result, to understand extra-mixing as a function of mass at low-metallicity using the MW alone is difficult.

Luckily, the LMC and SMC host a large population of young, higher-mass ($>$ 1.1 \Msun), metal-poor stars that can be used to refine the extra-mixing prescription as a function of mass.
The LMC intermediate-age globular clusters have been used as a stellar laboratory for extra-mixing \citep[e.g.,][]{Lederer2009}, but it is not until recently that a large number of field RGB stars were observed with all-sky spectroscopic surveys such as APOGEE \citep{Majewski2017}, SDSS-V MWM \citep{Kollmeier2026}, and GALAH \citep{DeSilva2015}.
These surveys provided us with detailed C and N abundances available to conduct extra-mixing studies with LMC field stars, as we did in the MW. 

However, a key caveat in using the MCs to study extra mixing as a function of stellar mass along the giant branch is that the birth abundance of [C/N] is not directly known. 
Lower RGB stars that have not yet undergone first dredge-up are intrinsically faint and therefore difficult to observe in the MCs. 
As a result, the initial [C/N] abundance, which sets the baseline for interpreting post-dredge-up measurements, remains poorly constrained.
This uncertainty is important because age determinations based on [C/N] rely on correcting the change in [C/N] induced during first dredge-up as a function of stellar mass and metallicity. 
The observable quantity is not the absolute [C/N], but rather its depletion relative to the birth composition. 
Consequently, uncertainties in the initial abundance propagate directly into uncertainties in inferred stellar masses and ages. 
It is therefore essential to develop independent constraints on the birth [C/N] distribution in the MCs before more observations are available.

Although the physical mechanism responsible for extra mixing remains uncertain, empirical evidence places limits on its depth. 
For example, \citet{Gratton2000} found no trend of [O/Fe] with $\log g$ among metal-poor stars ($-2 < \mathrm{[Fe/H]} < -1$), implying that extra mixing does not reach layers where ON cycling occurs. 
Under this constraint, [(C+N)/Fe] should remain approximately conserved even in stars that have experienced extra mixing, allowing it to serve as a tracer of the birth composition.
In addition, [Fe/H] provides an alternative proxy for the birth chemical mixture. 
In the MW, [Fe/H] is strongly correlated with birth [C/N] \citep{Roberts2024}. 
Leveraging this relationship offers a practical means of accounting for the initial abundance mixture in the MCs, particularly in cases where pre-dredge-up stars are too faint to observe directly.

In this paper, we test the extra-mixing prescription in \cite{Shetrone2019} and thermohaline mixing models in \cite{Lagarde2012} using the LMC and SMC field stars.
We describe the mass, age, and abundance data from the MCs, our comparison sample from the MW, and how the extra-mixing correction is applied in Section~\ref{sec:data}, and we describe our comparison results in Section~\ref{sec:results}.
Finally, we summarize our findings and future paths forward in Section~\ref{sec:conc}.

\section{Data \& Methods}\label{sec:data}
\subsection{Observational Data}
The full sample of $\sim$6000 LMC and $\sim$2000 SMC RGB stars were observed by APOGEE \citep{Majewski2017} with the target selection and membership determination described in \citet{Nidever2020}.
Chemical abundances were derived from APOGEE DR17 \citep{abdurrouf2022}, with spectra reduced using the APOGEE data reduction pipeline \citep{Nidever2015} and stellar parameters and abundances determined using ASPCAP \citep{garcia2016}.

\begin{figure*}[ht!]
\includegraphics[width=\textwidth]{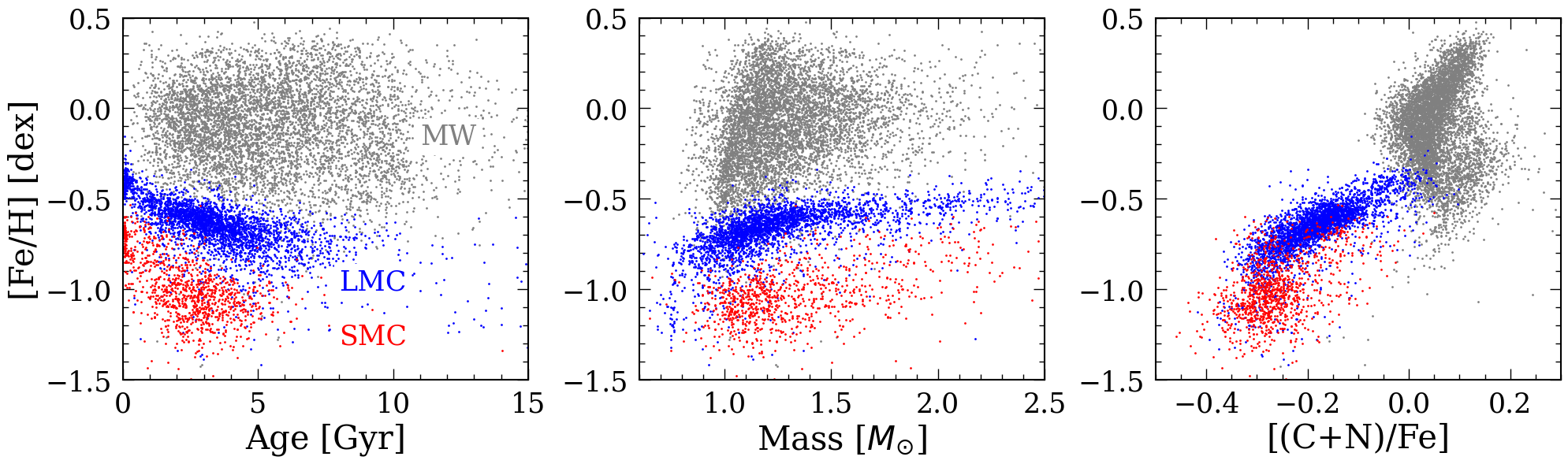}
\caption{The age-metallicity (left), mass-metallicity (middle), and [(C+N)/Fe]-[Fe/H] relations for the APOKASC--3 sample \citep{Pinsonneault2025} in gray, the LMC sample in blue \citep{abdurrouf2022, Povick2024}, and the SMC sample in red \citep{abdurrouf2022, Povick2024}.
The LMC and SMC harbor massive, metal-poor stars that can be used to understand and test extra-mixing correction as a function of mass at [Fe/H]$< -$0.5.
The strong correlation between [Fe/H] and [(C+N)/Fe] suggests [Fe/H] should be a good proxy for the birth mixture for stars in the MW and the MCs.
The two sequences observed in [(C+N)/Fe]–[Fe/H] space in the MW correspond to the high- and low-$\alpha$ disk populations. 
At a fixed [Fe/H], stars belonging to the high-$\alpha$ disk exhibit systematically higher [(C+N)/Fe] than those in the low-$\alpha$ disk.}
\label{fig:1}
\end{figure*}

Stellar ages and masses were derived following the method in \cite{Povick2024} using the observed APOGEE spectroscopic parameters ($T_{\rm eff}$, $\log{g}$, {[Fe/H]}, and {[$\alpha$/Fe]}), multi-band photometry from Gaia \citep{gaia, Gaiadr3} and 2MASS \citep{2mass}, and PARSEC isochrones \citep{parsec, parsec2, parsec3}. 
Distances for the stars were found using the red clump LMC model from SMASH \citep{Choi2018a, Choi2018b} and assuming that the RGB stars lie close to the disk plane.
For each star, model photometry and \logg\ were interpolated from the isochrones as functions of effective temperature, [Fe/H], [$\alpha$/Fe], and a trial age. 
Since the isochrones are based on scaled-solar models, the [Fe/H] are adjusted with the measured [$\alpha$/Fe] following \cite{Salaris1993} with updated coefficients from \cite{Povick2024}.
The trial age was then adjusted until the model photometry and \logg\ best matched the observed values. 
Validation with the APOKASC--2 \citep{Pinsonneault2018} asteroseismic sample (with accurate masses and ages) shows a 20\% error for our age estimates.

To construct the sample used in our paper, we selected stars with no [Fe/H], [C/Fe], or [N/Fe] ASPCAP flags, and with uncertainties in all three abundances $<$ 0.05 dex.
Note that the C and N measurement uncertainties reported in ASPCAP are likely too small \citep{Pinsonneault2025, Cao2026}, as a result, we only use the uncertainties as a guide to exclude outliers.
We further restrict the sample to stars with isochrone masses below 100 \Msun\ and mass uncertainties below 0.5 \Msun\ to eliminate implausible mass estimates.
Known stellar binaries with APOGEE RV scatter greater than 1 km s$^{-1}$ are also removed. The final sample contains $\sim$3200 LMC stars and $\sim$1100 SMC stars, with most rejected stars having masses below 1 \Msun\ and are excluded because of their large mass uncertainties.

To compare with stars in the MW, we use the APOKASC--3 asteroseismic catalog, with asteroseismic masses and \logg\ measured from Kepler light curves \citep{kepler} and abundances measured from APOGEE DR17 \citep{Pinsonneault2025}.
As the MC sample is restricted to upper RGB stars by magnitude limits, we select only RGB stars from APOKASC--3 for consistency and apply identical quality cuts to those used for the LMC and SMC.
These selections left us with $\sim$6000 stars.

The final sample is shown in \autoref{fig:1}, where the left and middle plots show the [Fe/H]-age and [Fe/H]-mass relation, respectively.
It is clear that the MCs host metal-poor stars of all masses.
The right plot shows the [Fe/H]-[(C+N)/Fe] relation.
The two sequences observed in [(C+N)/Fe]–[Fe/H] space in the MW correspond to the high- and low-$\alpha$ disk populations. 
At a fixed [Fe/H], stars belonging to the high-$\alpha$ disk exhibit systematically higher [(C+N)/Fe] than those in the low-$\alpha$ disk.
Similar to the MW, there exists a strong relation between [Fe/H] and [(C+N)/Fe] for the MCs, suggesting [Fe/H] can be used as a proxy for the birth [C/N] abundance, meaning stars with a similar [Fe/H] can be assumed to have a similar birth [C/N].

\begin{figure*}[ht!]
\includegraphics[width=\textwidth]{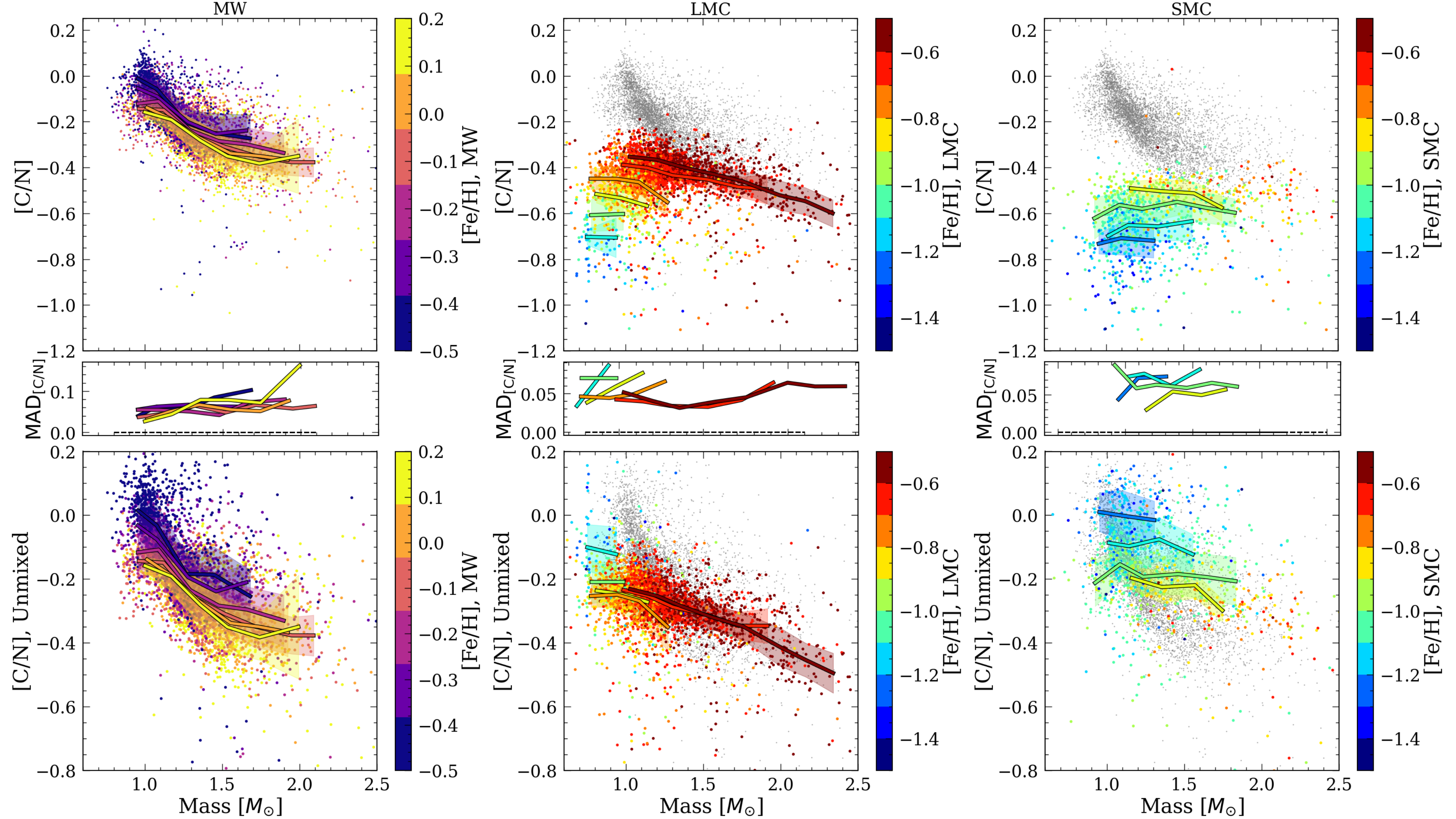}
\caption{The top row shows the observed [C/N] and the bottom row shows the [C/N] after correcting for extra-mixing following \cite{Shetrone2019}.
The points are colored by their metallicity.
The gray points in the middle and right columns show the [C/N]-mass relation for the MW (same as the corresponding left column) for better comparison.
The colored lines and shaded areas show the running median and median absolute deviation (MAD) for different mono-metallicity populations, respectively.
The subplots under each figure in the top row show the MAD as a function of mass for better visualization. 
The metal-poor, high-mass stars in the LMC and SMC hint that extra-mixing might not be effective for stars more massive than $\sim$1.8 \Msun\ at [Fe/H] $\sim -$0.7, where the running median of the lower metallicity line crosses that of the higher metallicity line in the LMC.
The lines also cross for the MW at $\sim$0.1 dex, respectively, most likely due to the increase in scatter around the line.
\label{fig:2}}
\end{figure*}

\subsection{Extra-Mixing Corrections}\label{subsec:em}
We test two extra-mixing corrections in this work, one empirically calibrated using MW high-$\alpha$ stars \citep{Shetrone2019}, and one based on stellar evolution models that include thermohaline mixing prescriptions \citep{Lagarde2012}.

To implement the empirical correction, we follow the values listed in Table 2 of \cite{Shetrone2019}.
We interpolated between metallicity and the change in [C/N] for extra-mixing and the additional change as a function of \logg\ below [Fe/H]$< -$1 using a linear interpolator. 
All stars with [Fe/H]$< -$1.6 were assumed to follow the same extra-mixing prescription. 

For the theoretical models presented in \cite{Lagarde2012}, we converted the initial total metallicity, $Z$, and the C and N elemental mass fractions ($X_{\rm C}$ and $X_{\rm N}$) into abundance ratios using solar abundances from \cite{Asplund2005}, to maintain consistency with both \cite{Lagarde2012} and the APOGEE DR17 abundance pipeline \citep[ASPCAP;][]{garcia2016}.

\section{Results} \label{sec:results}
\subsection{The Observed [C/N]-Mass Relations at Mono-Birth Mixtures}
In \autoref{fig:2}, we plot the overall mass-[C/N] relation for the three samples, with the observed [C/N] on the top row.
The points are colored by their metallicity values.
For the MW sample, we select only stars with \logg\ $<$ 3 to exclude any that have not gone through the first dredge-up to be consistent with the MC samples.
The colored lines and shaded areas are the running median and median absolute deviation (MAD), respectively, of [C/N] in mono-metallicity (0.1 dex in width) populations.
The MAD is also plotted separately in the subplots in each figure in the top row for better visualization. 
The medians are taken with a width of 0.5 \Msun\ and a sliding window of 0.2 \Msun. 
We require more than 15 data points in each mass bin for a valid running median calculation.

The MW sample serves as a nice reference case where the effect of extra-mixing is weak ([Fe/H] $> -$0.4).
For each metallicity bin, [C/N] decreases with increasing mass or decreasing age, as expected.
Since the effect of metallicity on the change of [C/N] during the first dredge-up is smaller than that of the birth mixture \citep[see Figures 9--11 in][]{Roberts2024}, the metal-rich RGB stars, on average, have a lower [C/N] compared to those that are metal-poor. 
At the high-mass end, the metal-rich line ([Fe/H]$\sim$0.1) appears to reverse and have higher [C/N]. 
However, this population is poorly sampled in the MW, as evidenced by the factor of 2$\times$ increase in the MAD, and thus this is likely an artifact rather than a real trend.

Extra-mixing can be seen in MW RGB stars between $-$0.4 dex and $-$0.5 dex as the median [C/N]-mass relation for this metallicity range lies under that where the stars are slightly metal-rich (top left plot).
As mentioned above, since the birth [C/N] increases towards lower [Fe/H] in the MW, the median [C/N] at the same mass should be higher for lower metallicity stars without extra-mixing.

For the LMC and SMC, obtaining spectra for stars with \logg\ $>$ 1.5 is difficult as they are faint. 
This means that we do not have access to the birth [C/N] mixture for stars in the MCs, which require spectra for subgiants that have not gone through the first dredge-up (about \logg\ $\sim$ 3.5). 
Future work can build chemical evolution or data-driven models to estimate the birth [C/N] mixture, but this is out of the scope of this paper. 
However, if we assume metallicity is also a good indicator of the birth [C/N] mixture in the MCs as a strong correlation between [Fe/H] and [(C+N)/Fe] exists (See \autoref{fig:1}), we can still perform the same analysis used for the MW to the MCs.

\begin{figure*}[ht!]
\includegraphics[width=\textwidth]{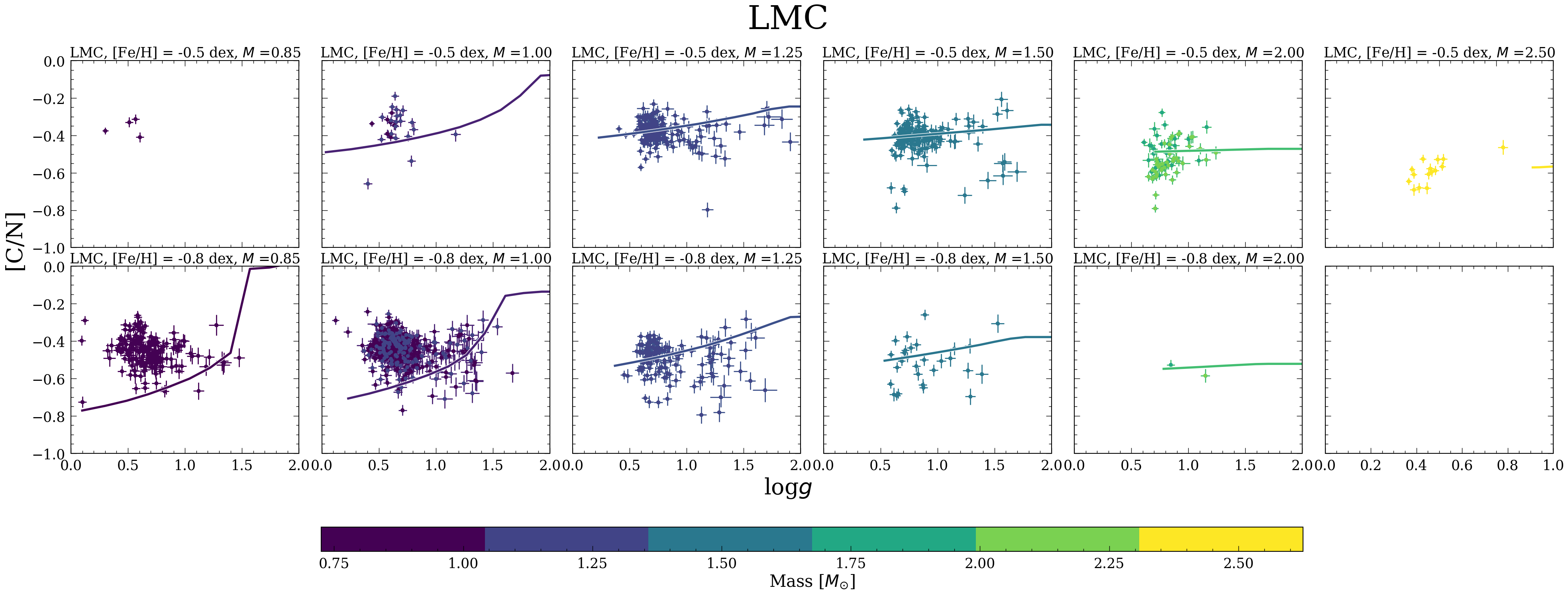}
\includegraphics[width=\textwidth]{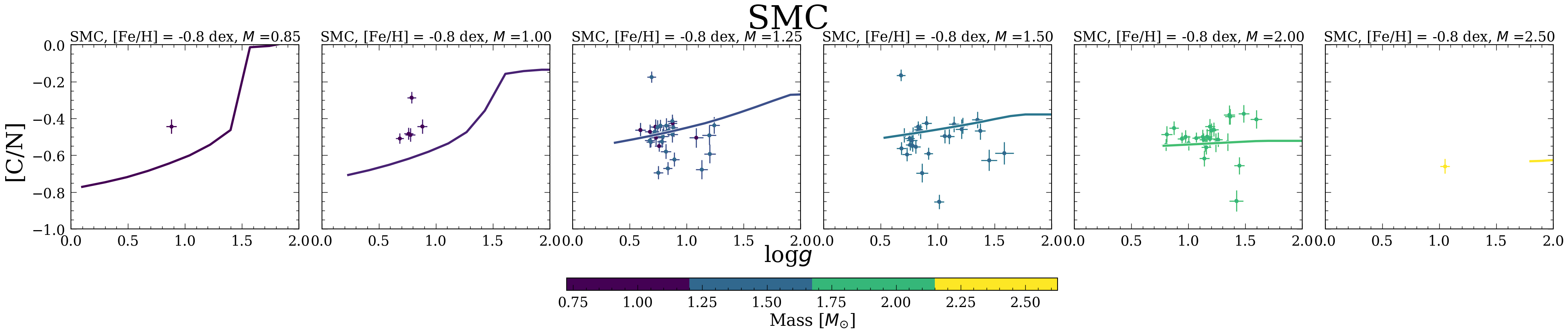}
\caption{Thermohaline + rotation mixing models from \cite{Lagarde2012} compared with the data in the LMC and SMC at different masses.
Each row shows stars in different metallicity bins and each column shows those in different mass bins.
The models are shifted up by 0.4 dex in [C/N] to better match the observations.
The models over-predict the \logg\ dependence of extra-mixing, as the data does not show a strong relation between \logg\ and [C/N], as also seen in the MW \citep{Shetrone2019}. 
An upturn of [C/N] is seen in the LMC at low \logg\ that could be caused by systematics in the abundance measurements \citep{Sit2024}. 
\label{fig:5}}
\end{figure*}

In contrast to the MW sample, the LMC and SMC populations exhibit systematically lower [C/N] at lower metallicity. 
This trend is unlikely to be driven by differences in the birth abundances. 
In the MCs, the strong age–metallicity relation implies that metal-poor stars are typically older. 
If birth [C/N] follows the common expectation of being higher at lower metallicity, then these old, low-mass, metal-poor stars should have formed with relatively elevated initial [C/N].
Moreover, because post–first-dredge-up [C/N] increases with decreasing stellar mass, one would expect low-mass, low-metallicity stars to display higher [C/N] overall. 
Instead, we observe the opposite behavior, strongly suggesting that additional mixing processes must be reducing the surface [C/N] in these stars.

Finally, in each metallicity range, [C/N] should decrease with increasing mass, as expected and also seen in the MW. 
For the LMC and SMC, this is mostly true down to [Fe/H] of $-$1.
However, the expected trend stagnates and reverses in the LMC at $\sim -$0.7 dex and in the SMC at $\sim -$1 dex. 
Unlike the opposite trend at higher metallicity in the MW, the MADs between the two mono-metallicity populations in the LMC are similar where the reversal occurs (between 1.5--2 \Msun).
If metallicity is an accurate representation of the birth [C/N] mixture, the reversal either represents that higher mass stars dredge up less [C/N] at lower metallicity compared to higher
metallicity, or that extra-mixing is less effective at higher masses.
Even though the former is possible, it is more likely the latter that is responsible for the reversal, as it is predicted by theory that thermohaline mixing stops being effective at a mass around 1.5--2 \Msun \citep[e.g.,][]{Cantiello2010, Charbonnel2010, Lagarde2012}. 
This is the first time observational evidence is presented for this mass threshold, and it is only made possible thanks to the recent large spectroscopic surveys.
More data are needed for the SMC to confirm this trend.

\begin{figure*}[ht!]
\includegraphics[width=\textwidth]{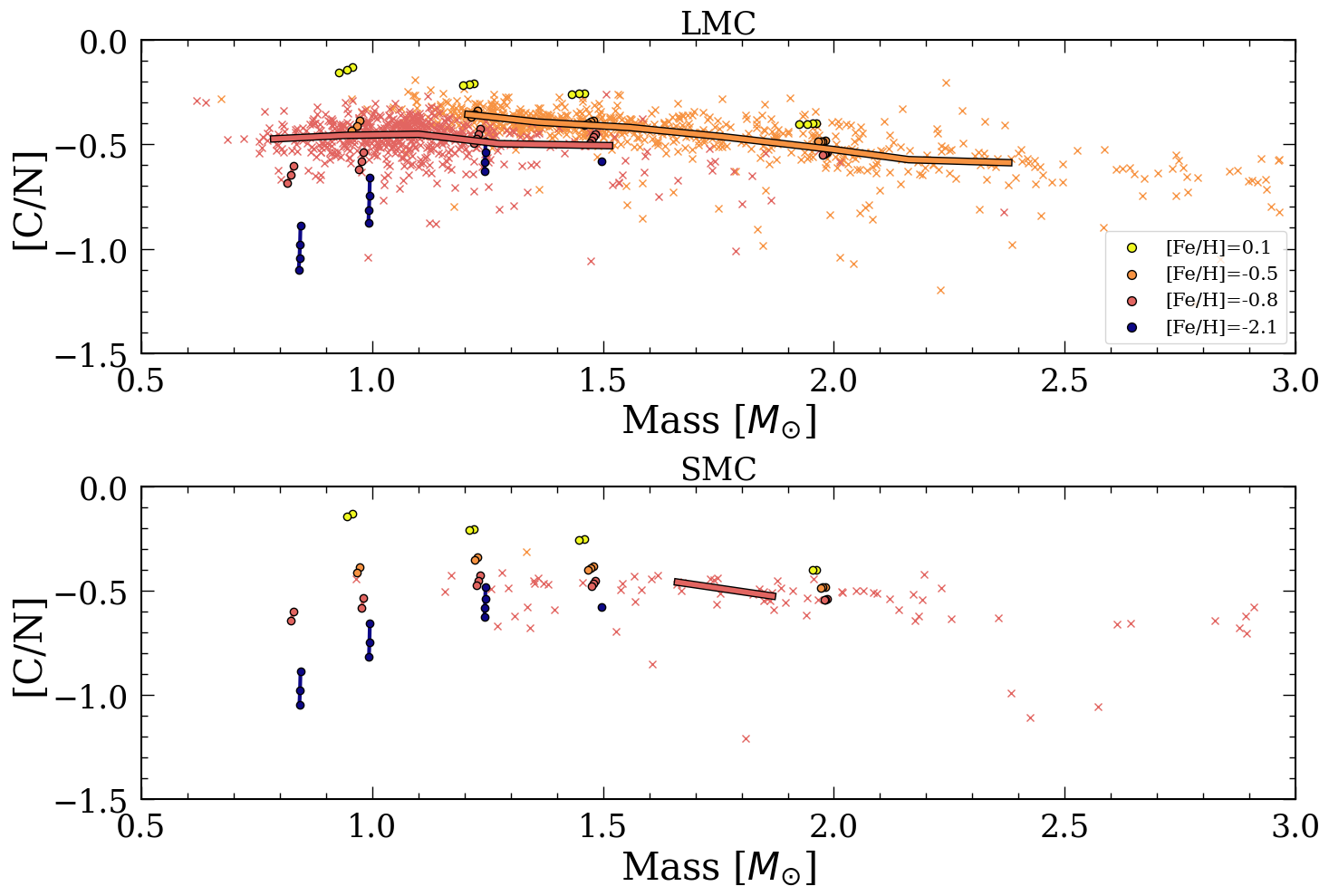}
\caption{Models from \cite{Lagarde2012} comparing with the LMC (top) and SMC (bottom) data. 
The initial $Z$ and mass fractions are converted to abundance ratios using the solar composition from \cite{Asplund2005}. 
The masses for the models are taken to be the initial mass. 
The models are plotted in black outlined points, colored by their initial metallicity.
Stars with the same initial mass and metallicity are connected by lines for better visualization.
The scatter for each metallicity at the same mass for the models is caused by the \logg\ selection, where thermohaline mixing continues to affect the [C/N] ratio in a \logg\ dependent way.
We only selected RGB stars with \logg\ between 0.6 and 1.2 for the LMC and 0.8 and 1.2 for the SMC to match the observed range.
The range of [C/N] for each model at a metallicity and mass can also be seen in \autoref{fig:5}.
We shift the model up by 0.4 dex to better match the data.
For the observations, we select the same \logg\ range for consistency.
The metallicity range is selected to be 0.1 dex around the model metallicity.
The data is plotted in crosses, colored by their [Fe/H], and the lines show the running median for better visualization.
The models fit well at [Fe/H] = $-0.5$ for the LMC but likely over-predict the effect of extra-mixing at $-0.8$ dex at lower masses ($\sim$1.2 \Msun). 
\label{fig:4}}
\end{figure*}

\subsection{Empirical Extra-mixing Corrections}
The [C/N] after correcting for extra-mixing using the empirical relation described in \cite{Shetrone2019} is shown in the bottom row of \autoref{fig:2}. 
The correction is mostly able to account for extra-mixing, as the MW and MCs now exhibit trends where the more metal-rich stars (more massive and younger) have a lower median unmixed [C/N] value below [Fe/H] of $-$0.5 dex, where extra-mixing is in effect.
However, the correction is not perfect, especially for the LMC.
This could either be caused by the difference in birth [C/N] mixtures or mass-dependent extra-mixing effects, as the correction in \cite{Shetrone2019} only used high-$\alpha$ stars in the MW to determine the empirical correction, which only includes stars that are $\sim$ 1 \Msun. 

The \logg\ range for the stars in the MCs is small (0.6--1.2), and [C/N] should not change with \logg\ in this regime according to data in the MW \citep{Shetrone2019}. 
This can indeed be seen in \autoref{fig:5}, where there is no strong \logg\ dependence for stars in the LMC and SMC at all masses for [Fe/H] of $-$0.5 and $-$0.8 dex.
This also holds true for the full metallicity range. 

\subsection{Theoretical Thermohaline Models}
We compare observational results in the LMC and SMC with models presented in \citep{Lagarde2012}. 
For each modeled [Fe/H], we select observed [Fe/H] measurements that are within 0.1 dex of the metallicity of the model.
We also shift the models up by 0.4 dex in [C/N] to match the abundance measurements.
The shift is likely due to the lower birth mixture in the MCs compared to that of the Sun (See \autoref{fig:1} right plot), as the models assumes solar mixture at the start of first dredge-up.

First, we compare predicted [C/N] from the \cite{Lagarde2012} thermohaline mixing + rotation models at [Fe/H] = $-0.5$ and [Fe/H] = $-0.8$ for different masses as a function of \logg. 
To do so, we select observations from the LMC and SMC with [Fe/H] within 0.1 dex and masses within 0.125 \Msun\ of the model metallicity and masses, respectively.

\autoref{fig:5} shows the comparison results for the LMC (top) and SMC (bottom).
The models fit remarkably well for the average [C/N] at mass $>$ 1.25 \Msun, where we see higher mass stars exhibit lower [C/N] compared to lower mass stars.
However, the model over-predicts the amount of mixing for lower mass stars, where \cite{Lagarde2012} models predict stars with mass $< $ 1 \Msun\ should end up with a lower [C/N] at the tip of the RGB compared to higher mass stars due to extra-mixing, where the data in the MCs show continue increase of [C/N] towards lower mass stars at [Fe/H] of $-$0.5 and $-$0.8 dex.
The [C/N]-mass relation for the SMC shown in \autoref{fig:2} suggests this should not happen until [Fe/H] $< -$1, where [C/N] downturns at mass $< \sim$ 1.25 \Msun.
This could potentially be explained by a [Fe/H] offset between the models and observation or systematics in abundance measurements and isochrone fitting; however, additional model grids with finer sampling in metallicity are required to robustly test and confirm this hypothesis.
There also exists a peculiar upturn in [C/N] at low \logg\ for masses below 1 \Msun\ in the LMC data.
This could be due to the \logg\ bias in APOGEE DR17 abundances \citep{Sit2024}, where systematic trends in abundances with \logg\ exist that are likely not physical.
Future work should correct for the bias in C and N as a function of \logg\ and re-evaluate the comparison. 

Next, we compare the observed [C/N]-mass relations with the models. 
To do so, we select the models with \logg\ between 0.6 and 1.2 to mimic the range in the LMC and SMC.
For the observations, we also perform the same \logg\ selection for consistency.
\autoref{fig:4} shows the comparison results, where the crosses and lines are the observations and running median, respectively.
The model predictions are shown as outlined points, with the scatter caused by the range in \logg.

At [Fe/H] = $-0.5$, the model agrees well with the data in the LMC.
At [Fe/H] = $-0.8$, the data only agrees well with the model in SMC stars, which have masses between 1.5 to 2 Msun. 
However, it over-predicts the extra-mixing for stars in the LMC as [C/N] stays almost flat for the data $<$ 1.2 \Msun, which is also seen in \autoref{fig:5}.
Again, at [Fe/H] = $-0.8$, the model predicts a downturn in [C/N] at low mass, which is not seen in the data until [Fe/H] = $-1$ (see \autoref{fig:2} top right plot for the SMC).

\section{Discussion \& Future Work}\label{sec:discussion}
This work highlights the growing potential for advancing our understanding of stellar interior physics and calibrating the [C/N]–age relation in the low-metallicity regime by leveraging current and upcoming large-scale spectroscopic and photometric surveys. 
In this context, the MCs represent the next-largest resolved stellar systems beyond the MW, uniquely providing access to a broad range of stellar masses and ages at subsolar metallicities. 
Their extended star formation histories make them a powerful laboratory for testing and refining models of stellar mixing down to [Fe/H] $\sim-0.7$, where constraints from the MW alone are limited by strong age–mass–metallicity degeneracies.

Moving forward, several developments will be critical. 
On the theoretical side, stellar evolution models with finer metallicity spacing and improved prescriptions for thermohaline and related mixing processes are needed to enable more precise comparisons with observations. 
In particular, model grids that densely sample metallicity between $-0.5$ and $-2$ dex would allow us to better quantify the onset and mass dependence of extra mixing.

On the observational side, extending [C/N] measurements to stars with higher surface gravities in the MCs would provide leverage on evolutionary stage and reduce systematic uncertainties associated with post-first dredge-up evolution. 
Deeper spectroscopic observations targeting fainter populations will also expand the accessible mass range and improve statistical constraints.
Finally, further analysis of MC star clusters, where ages and metallicities are independently constrained, offers a crucial pathway for validating and refining mass estimates. 
Such benchmarks will be essential for anchoring empirical [C/N] calibrations and for testing stellar evolution models in regimes not accessible within the MW.

Together, these efforts will pave the way toward robust [C/N]-based ages for individual stars in external galaxies and enable the use of resolved stellar populations as precision probes of galaxy formation histories.

\section{Conclusion and Future Work}\label{sec:conc}
In this paper, we showcase the potential of understanding and correcting for extra-mixing as a function of mass at [Fe/H] $< -$ 0.5. 
We test the empirical correction from \cite{Shetrone2019} and theoretical stellar models with thermohaline and rotation mixing prescription from \cite{Lagarde2012}. 
Overall, the empirical calibration performs well at lower mass $<$ 1.25 \Msun\ and the theoretical models are able to capture the mass-dependence well down to 1.25 \Msun. 
Using the LMC and SMC data, we are able to identify the mass-dependency and the mass threshold ($\sim$1.8 \Msun\ at [Fe/H] $\sim -$0.7) where extra-mixing stops being effective for the first time. 
However, since the empirical correction from \cite{Shetrone2019} is calibrated on the metal-poor stars in the MW, unsurprisingly, it cannot be applied to higher masses, such as those in dwarf galaxies, as the strength of extra-mixing is a function of mass. 
On the other hand, the theoretical models work well to capture the mass dependency and are able to predict the general behavior at different masses for [Fe/H] $> -$0.8 and mass $>$ 1.25 \Msun, but over-predicts the effect of mixing at $<$ 1.25 \Msun.
The \logg\ range of the stars in the MCs is of 0.6-1.2, in which \cite{Shetrone2019} predicts no strong correlation between [C/N] and \logg, agreeing with the data, and the models from \cite{Lagarde2012} over-predicts the \logg\ dependency in this regime. 

\begin{acknowledgments}
Y.L. wants to thank the stars group at OSU for the helpful discussion. 
D.L.N. acknowledges support from NSF grants AST 1908331 and 2408159.
J.D.R., M.H.P., and J.A.J. acknowledge support from NASA ADAP grant 80NSSC24K0637. J.A.J. also acknowledges support from NSF grant AST- 2307621
This work has made use of data from the European Space Agency (ESA) mission Gaia,\footnote{\url{https://www.cosmos.esa.int/gaia}} processed by the Gaia Data Processing and Analysis Consortium (DPAC).\footnote{\url{https://www.cosmos.esa.int/web/gaia/dpac/consortium}} 
Funding for the DPAC has been provided by national institutions, in particular the institutions participating in the Gaia Multilateral Agreement.
This research also made use of public auxiliary data provided by ESA/Gaia/DPAC/CU5 and prepared by Carine Babusiaux. 
This research has also made use of NASA's Astrophysics Data System.

Portions of the writing and editing in this paper were assisted with the help of the ChatGPT large language model (OpenAI, 2025). The tool was used to improve clarity, phrasing, grammar, and structure of draft text, while all scientific content, interpretations, and conclusions remain entirely the responsibility of the authors.

\end{acknowledgments}

%

\vspace{5mm}
\facilities{Gaia \citep{gaia, Gaiadr3}, 2MASS \citep{2mass}, APOGEE--2S \citep{Wilson2019}, Kepler \citep{kepler}}


\software{astropy \citep{astropy:2013, astropy:2018, astropy2022}, Matplotlib \citep{matplotlib}, NumPy \citep{Numpy}, Pandas \citep{pandas}, ChatGPT \citep{openai2025chatgpt}}






\bibliography{sample631}{}
\bibliographystyle{aasjournal}



\end{document}